\documentclass[conference]{IEEEtran}
\IEEEoverridecommandlockouts
\usepackage{cite}
\usepackage{amsmath,amssymb,amsfonts}
\usepackage{algorithmic}
\usepackage{algorithm}
\usepackage{graphicx}
\usepackage{textcomp}
\usepackage{xcolor}
\usepackage{url}
\usepackage{graphicx}
\usepackage{csquotes}
\usepackage[nolist]{acronym}
\usepackage{subfigure}
\usepackage{stfloats}
\def\BibTeX{{\rm B\kern-.05em{\sc i\kern-.025em b}\kern-.08em
		T\kern-.1667em\lower.7ex\hbox{E}\kern-.125emX}}
\begin{document}
	
	\title{GLDPC-PC Codes for MIMO Systems with \\ Iterative Detection and Decoding
	}
	
	\author{
		\IEEEauthorblockN{Binghui Shi, Yongpeng Wu, Yin Xu, Xiqi Gao, Xiaohu You and Wenjun Zhang}
		\thanks{B. Shi, Y. Wu, Y. Xu, and W. Zhang are with the Department of Electronic Engineering, Shanghai Jiao Tong University, Shanghai 200240, China. (e-mail: zeppoe@sjtu.edu.cn; yongpeng.wu@sjtu.edu.cn; xuyin@sjtu.edu.cn; zhangwenjun@sjtu.edu.cn). \textit{(Corresponding author: Yongpeng Wu.)}}
		\thanks{X. Gao and X. You are with the National Mobile Communications Research Laboratory, Southeast University, Nanjing 210096, China. (e-mail: xqgao@seu.edu.cn; xhyu@seu.edu.cn).}
	}
	
	\maketitle
	
	\begin{acronym}
		\acro{MIMO}{multiple-input multiple-output}
		\acro{LMMSE}{linear minimum mean square error}
		\acro{MSE}{mean square error}
		\acro{APP}{a-posteriori probability}
		\acro{AUB}{approximated union bound}
		\acro{AWGN}{additive white Gaussian noise}
		\acro{B-DMC}{binary-input discrete memoryless channel}
		\acro{BCJR}{Bahl, Cocke, Jelinek and Raviv}
		\acro{BEC}{binary erasure channel}
		\acro{BER}{bit error rate}
		\acro{biAWGN}{binary-input additive white Gaussian noise}
		\acro{BLER}{block error rate}
		\acro{bpcu}{bits per channel use}
		\acro{BPSK}{binary phase-shift keying}
		\acro{BP}{belief propagation}
		\acro{CN}{check node}
		\acro{CRC}{cyclic redundancy check}
		\acro{CSI}{channel state information}
		\acro{DE}{density evolution}
		\acro{DMC}{discrete memoryless channel}
		\acro{DMS}{discrete memoryless source}
		\acro{dRM}{dynamic RM}
		\acro{FER}{frame error rate}
		\acro{uFER}{undetected frame error rate}
		\acro{FHT}{fast Hadamard transform}
		\acro{GA}{Gaussian approximation}
		\acro{GF}{Galois field}
		\acro{GLDPC}{generalized low-density parity-check}
		\acro{GLDPC-PC}{GLDPC with polar component}
		\acro{HARQ}{hybrid automated repeat request}
		\acro{IDD}{iterative detection and decoding}
		\acro{i.i.d.}{independent and identically distributed}
		\acro{LDPC}{low-density parity-check}
		\acro{LHS}{left hand side}
		\acro{LLR}{log-likelihood ratio}
		\acro{MAP}{maximum-a-posteriori}
		\acro{MC}{Monte Carlo}
		\acro{MLC}{multilevel coding}
		\acro{MLPC}{multilevel polar coding}
		\acro{ML}{maximum-likelihood}
		\acro{MMSE}{minimum mean-square error}
		\acro{NPI}{noise-plus-interference}
		\acro{PAC}{polarization-adjusted convolutional}
		\acro{PAT}{pilot-assisted transmission}
		\acro{PCM}{polar-coded modulation}
		\acro{PDF}{probability density function}
		\acro{PE}{polar extension}
		\acro{PIC}{parallel interference cancellation}
		\acro{PMF}{probability mass function}
		\acro{PM}{path metric}
		\acro{PW}{polarization weight}
		\acro{QAM}{quadrature amplitude modulation}
		\acro{QC}{quasi-cyclic}
		\acro{QPSK}{quadrature phase-shift keying}
		\acro{QUP}{quasi-uniform puncturing}
		\acro{RCU}{random-coding union}
		\acro{RHS}{right hand side}
		\acro{RM}{Reed-Muller}
		\acro{RQUP}{reversal quasi-uniform puncturing}
		\acro{RV}{random variable}
		\acro{SC}{successive cancellation}
		\acro{SCL}{successive cancellation list}
		\acro{SO-SCL}{soft-output SCL}
		\acro{SISO}{soft-input soft-output}
		\acro{SPC}{single parity-check}
		\acro{SNR}{signal-to-noise ratio}
		\acro{VN}{variable node}
		\acro{5G}{the $5$-th generation wireless system}
	\end{acronym}

	\begin{abstract}
		In this work, we propose the integration of GLDPC codes with short polar-like component codes, termed GLDPC codes with polar component codes (GLDPC-PC). This approach leverages the good distance properties of polar-like codes and mitigates their high decoding latency in long block lengths. A recently proposed soft-input soft-output decoder for polar-like codes enables effective iterative belief propagation decoding for GLDPC-PC, ensuring a low error floor under additive white Gaussian noise channels. Simulation results demonstrate that GLDPC-PC codes achieve significant performance improvements in multiple-input multiple-output systems with iterative detection and decoding (IDD). The proposed GLDPC-PC codes and the IDD scheme can be applied to various scenarios.
	\end{abstract}
	
	
	\section{Introduction}
	
	\Ac{GLDPC}~\cite{gldpc-init} code generalizes \ac{LDPC} code~\cite{ldpc} by replacing the \ac{SPC} constraints with more general constraints based on stronger component codes. The \ac{BP} decoding of \ac{GLDPC} codes iteratively pass extrinsic information between the \acp{VN} and \acp{CN} in the Tanner graph where every \ac{CN} represents a component code. The \ac{SISO} decoder for the component code is necessary for \ac{BP} decoding. One of the major obstacles in the development of \ac{GLDPC} codes is the lack of low-complexity \ac{SISO} decoders.
	
	The optimal \ac{SISO} decoding, called \ac{BCJR} algorithm, is proposed in~\cite{bcjr}. However, its computational complexity grows exponentially for linear block codes in the number of parity checks. \cite{pyndiah} proposes an approximation based on list decoding but lacks precision and requires optimization of a weight factor and a saturation value. Whereas acquiring a \ac{SISO} decoder with both satisfying complexity and precision for any linear block code is difficult, it can be achieved for specific code.
	
	Polar-like codes represent a wide range of modifications of polar codes, whose codebook is obtained by performing a linear transformation on polar codebook~\cite{polar}. For example, \ac{CRC} concatenated polar codes~\cite{crc-polar} exhibit excellent performance under \ac{SCL} decoding in short block length. \Ac{PAC} codes~\cite{pac} have been proven to possess better distance properties~\cite{minimal}. A recent work~\cite{yuan} introduces \ac{SO-SCL} to extract the bitwise soft information for polar-like codes based on the codebook probability. 
	
	\Ac{MIMO} technology is widely used in current communication systems to achieve high data rates. A near-capacity receiver scheme is \ac{IDD}~\cite{idd}, which entails the \ac{SISO} decoder for the channel coding. For the purpose of low complexity, \ac{MMSE} \ac{PIC}~\cite{mmse-init} \ac{MIMO} detection is proposed. Its sub-optimal performance further emphasizes the necessity of \ac{IDD} scheme.
	
	We propose a novel coding scheme, termed as \ac{GLDPC-PC} codes, by integrating the concept of \ac{GLDPC} and \ac{SISO} decoding with polar-like codes. This approach utilizes the superior distance properties of polar-like codes and avoids their high decoding complexity at longer block lengths, as the \ac{SC}-based decoding for these codes requires a super-linear complexity. In the \ac{BP} decoding for \ac{GLDPC-PC} codes, the \acp{CN} are updated by \ac{SO-SCL}. Furthermore, we analyze the performance of \ac{GLDPC-PC} code in \ac{AWGN} channel and \ac{MIMO} \ac{IDD} systems with this \ac{BP} decoder. Although we focus on the point-to-point MIMO systems, the proposed \ac{GLDPC-PC} codes and the \ac{IDD} approach can be applied to various scenarios, e.g., multiuser MIMO and masssive MIMO systems.
	
	This paper is organized as follows. Section~\ref{section2} introduces the polar-like codes, their decoding and the basic concept of \ac{GLDPC}. Section~\ref{section3} introduces \ac{GLDPC-PC} code and its \ac{BP} decoding. The model of \ac{MIMO} \ac{IDD} system is given in Section~\ref{section4}. Section~\ref{section5} shows the simulation results and Section~\ref{section6} concludes the paper.
	
	\section{Preliminaries}
	\label{section2}
	
	\subsection{Notation}
	
	Vectors of length $N$ are denoted as $\boldsymbol{x}$ or $x^N =	(x_1, x_2, \cdots , x_N)$, where $x_i$ is its $i$-th entry. For natural numbers, we write $\left[a\right] = \left\{ i \in \mathbb{Z}\mid 1 \leq i \leq a \right\}$. For $x^N$, $x^a$ where $a\in\left[N\right]$ denotes the sub-vector $(x_1, x_2, \cdots , x_a)$. Uppercase letters, e.g., $X$, denote random variables while lowercase letters, e.g., $x$, denote their realizations. The probability distribution of $X$ evaluated at $x$ is written as $P_X(x)$. Uppercase bold letters, e.g., $\mathbf{X}$, denote matrices. $\log$ denotes natural logarithm without special statement.
	
	\subsection{Polar-like Codes}
	
	A binary polar-like code of block length $N$ and $K$ information bits is defined by a set $\mathcal{A} \subseteq \left[ N \right]$ with $\left| \mathcal{A} \right|=K$ and a transfer matrix $\mathbf{T}$ of size $N\times N$ over $\mathrm{GF}(2)$, where $N$ is a positive-integer power of 2 and $\mathcal{F} \triangleq \left[N\right] \backslash \mathcal{A}$. The encoding of polar-like codes is implemented by two transforms
	\begin{equation}
		c^N=u^N \mathbf{T} \mathbf{P}_n,
		\label{encodeing} 
	\end{equation}
	where $c^N$ is the codeword, $u^N$ is the input of the two transforms and $\mathbf{P}_n=\mathbb{F}^{\otimes n}$ is the $n$-th Kronecker power of binary Hadamard matrix $\mathbb{F}$ with $n=\log _2 N$. The $K$ information bits $s^K$ are placed in the subvector $u_{\mathcal{A}}$ of $u^N$ and $u_{\mathcal{F}}$ is set to 0. We demand $\mathbf{T}=\mathbf{I}_N+\mathbf{R}$, where $\mathbf{R}$ is a strictly upper triangular matrix and $\mathbf{R}_{j,i} = 1$ only if $j\in\mathcal{A}$ and $i\in\mathcal{F}$. Let $\mathcal{A}_i, i\in\mathcal{F}$ denote the indices of 1 in $i$-th column of $\mathbf{R}$. Obviously, if $j\in\mathcal{A}_i$, $j<i$ and $\mathcal{A}_i \subseteq \mathcal{A}$.

	The type of frozen constraints where the frozen bits are set to 0 statically is called static frozen bits, i.e., $\mathbf{T}=\mathbf{I}_N$. The other type of frozen constraints where the frozen bits $u_i$ are linear functions of $u^{i-1}$ is called dynamic frozen bits, for instance~\cite{pac}, 
	\begin{equation}
		\mathcal{A}_i=\left\{i-2,i-3,i-5,i-6\right\}\cap\mathcal{A}, i \in \mathcal{F}.
		\label{polarlike}
	\end{equation}
	This representation includes various modifications of polar codes, e.g., \ac{CRC}-concatenated polar codes~\cite{crc-polar},  and dynamic \ac{RM} codes~\cite{drm} and \ac{PAC} codes \cite{pac}. After $u^N$ is determined, the codeword $c_N$ is obtained by polar transform and then transmitted through \ac{B-DMC} with channel output $y^N$.
	
	
	\subsection{SC-based Decoding for Polar-like Codes}
	
	A common decoding method for polar codes is SC decoding, which can be introduced through the concept of binary decision trees. Every node of stage $i$ has two edges to the next stage and these two edges corresponds to the decision for $u_i = 0$ or $u_i = 1$. At a node of stage $i$, the \ac{SC} decoding performs a greedy decision based on the estimation $\hat{u}^{i-1}$ and obtains a decision of $u_i$ as
	\begin{equation}
		\hat{u}_i= \begin{cases}\bigoplus \limits_{j \in \mathcal{A}_i}
			u_{j}, & i \in \mathcal{F} \\ \underset{u \in\{0,1\}}{\arg \max} Q_{U_i \mid Y^N U^{i-1}}\left(u \mid y^N \hat{u}^{i-1}\right). & i \in \mathcal{A}\end{cases}
		\label{greedy}
	\end{equation}
	where $Q_{U_i \mid Y^N U^{i-1}}\left(u \mid y^N \hat{u}^{i-1}\right)$ denotes an auxiliary conditional \ac{PMF} induced by assuming uniform $U^N$ over $\{0, 1\}^N$~\cite{yuan}. \Ac{SC} decoding successively adopts this greedy decision from the root node to a leaf node. A path from the root node to a leaf node is called a decoding path and all the decoding paths form a set $\mathcal{U}$, specifically,
	\begin{equation}
		\mathcal{U} \triangleq\left\{u^N \in\{0,1\}^N\mid u_i=f_i\left(u^{i-1}\right), \forall i \in \mathcal{F}\right\} .
	\end{equation}
	
	\ac{SC} decoding only searches one path of the binary decision tree. However, \ac{SC} decoding has the ability to provide a \ac{PM} $Q_{U^i \mid Y^N}\left(\hat{u}^i \mid y^N\right)$ to weight different partial decoding paths $\hat{u}^{i}$. Therefore, it is possible to track more than one paths and avoid the greedy decision. \Ac{SCL} decoding implemented this idea via tracking $L$ decoding paths with largest \ac{PM}.
	
	At the nodes of stage $i$, \ac{SCL} decoding calculate the \ac{PM} of the $2L$ new paths from the existing $L$ paths and store $L$ paths with largest \ac{PM} in these $2L$ new paths. By the Bayes’ rule, 
	\begin{equation}
		\begin{aligned}
			& Q_{U^i \mid Y^N}\left(\hat{u}^i \mid y^N\right) \\
			& =Q_{U^{i-1} \mid Y^N}\left(\hat{u}^{i-1} \mid y^N\right) Q_{U_i \mid Y^N U^{i-1}}\left(\hat{u}_i \mid y^N \hat{u}^{i-1}\right),
		\end{aligned}
	\end{equation}
	where the term $Q_{U_i \mid Y^N U^{i-1}}\left(\hat{u}_i \mid y^N \hat{u}^{i-1}\right)$ can be computed efficiently by the standard SC decoding and $Q_{U^0 \mid Y^N}\left(\varnothing \mid y^N\right) \triangleq 1$ by definition. Let $\mathcal{L}$ denote the set of $L$ decoding paths when \ac{SCL} decoding terminates and obviously $\mathcal{L} \subseteq \mathcal{U}$. \cite{llr-scl} modifies the \ac{PM} to $-\log  Q_{U^i \mid Y^N}\left(\hat{u}^i \mid y^N\right)$ to change \ac{SCL} decoding into \ac{LLR} domain and adopts several approximations to reduce the computational complexity of \ac{SCL}.
	
	
	\section{GLDPC-PC and BP Decoding}
	\label{section3}
	
	\subsection{Code Structure}
	\label{pcgldpc}
	
	\Ac{GLDPC} code extends \ac{LDPC} code through replacing the \ac{SPC} of every row in parity check matrix by several simply component codes~\cite{replace}. \Ac{GLDPC} code consists of an adjacency matrix $\mathbf{\Gamma}$ of size $m\times N$ and $m$ component codes with parity check matrix $\mathbf{H}_k,k\in\left[m\right]$. These $m$ component codes can be identical or different. The parity check matrix $\mathbf{H}$ of \ac{GLDPC} is obtained via replacing the $j$-th 1 in $k$-th row of $\boldsymbol{\Gamma}$ by the $j$-th column of $\mathbf{H}_k$. Therefore, the number of column of $\mathbf{H}_k$ must be equal to the number of 1 in $k$-th row of $\mathbf{\Gamma}$.
	
	\Ac{GLDPC-PC} code refers to \ac{GLDPC} code adopting polar-like codes as component code. The consideration of \ac{GLDPC-PC} is to utilize the excellent performance and low-complexity \ac{SISO} decoder of polar-like codes in short block length and avoid the high computational complexity of polar-like codes in long block length. The parity check matrix $\mathbf{H}_k$ of polar-like component codes can be conveniently obtained from $\mathbf{P}_n$ and $\mathbf{T}$. After replacement, the overall parity check matrix $\mathbf{H}$ is obtained. Gauss elimination is performed to $\mathbf{H}$ to eliminate the linearly dependent rows and obtain the generator matrix for systematic encoding.
	
	\cite{gldpc} presents a \ac{GLDPC-PC} code which has significantly lower error floors. The \ac{GLDPC-PC} code adopts the (32,26) \ac{RM} code as component code and adopts adjacency matrix
	\begin{equation}
		\mathbf{\Gamma}_0=\left[\begin{array}{cccccc}
			\gamma^0 & \gamma^0 & \gamma^0 & \ldots & \gamma^0 & \gamma^0 \\
			\gamma^0 & \gamma^1 & \gamma^2 & \ldots & \gamma^{\tilde{n}-2} & \gamma^{\tilde{n}-1}
		\end{array}\right],
	\end{equation}
	where $\gamma$ is obtained by right rotation for 1 position of $\mathbf{I}_{\tilde{n}}$ and $\gamma^0 = \mathbf{I}_{\tilde{n}}$. $\mathbf{\Gamma}_0$ has \ac{QC} structure and $\tilde{n} = 32$ since the number of 1 in every row of $\mathbf{\Gamma}_0$ must be 32. The component codes of first $\tilde{n}$ rows are the initial (32,26) \ac{RM} code, and a random permutation is applied to the \ac{RM} code for the component codes of last $\tilde{n}$ rows. For most cases, a (1024, 643) \ac{GLDPC-PC} code is obtained after Gauss elimination.
	
	\subsection{SISO Decoding for Polar-like Codes}
	
	\Ac{BP} decoder for \ac{GLDPC} requires the \ac{SISO} decoder for its component codes. Therefore, the \ac{SISO} for polar-like codes is necessary for \ac{GLDPC-PC}.
	
	\Ac{SISO} decoding is a kind of decoding methods intended to minimize the bit error rate for a linear block code. A \ac{SISO} decoder treats the sum of channel observation in \ac{LLR} form $L_i$ and a-priori information in \ac{LLR} form $L_i^{\mathrm{A}}$ as input, where
	\begin{equation}
		L_i \triangleq \log\frac{p_{Y \mid C}\left(y_i \mid 0\right)}{p_{Y \mid C}\left(y_i \mid 1\right)},~L_i^{\mathrm{A}} \triangleq \log \frac{P_{C_i}(0)}{P_{C_i}(1)}, i\in\left[N\right].
	\end{equation}
	The output of it is \ac{APP} in \ac{LLR} form and extrinsic information in \ac{LLR} form, representing as $L_i^{\mathrm{APP}}$ and $L_i^{\mathrm{E}}$ respectively. To be specific
	\begin{equation}
		\begin{aligned}
			L^{\mathrm{APP}}_i & \triangleq \log \frac{P_{C_i \mid Y^N}\left(0 \mid y^N\right)}{P_{C_i \mid Y^N}\left(1 \mid y^N\right)} \\
			& =\log \frac{\sum_{c_i=0, c^N \in \mathcal{C}} P_{C^N \mid Y^N}\left(c^N \mid y^N\right)}{\sum_{c_i=1, c^N \in \mathcal{C}} P_{C^N \mid Y^N}\left(c^N \mid y^N\right)},
			\\L_i^{\mathrm{E}} &\triangleq L_i^{\mathrm{APP}} - L_i^{\mathrm{A}} - L_i,i\in\left[N\right],
		\end{aligned}
		\label{decoder}
	\end{equation}
	where $\mathcal{C}$ is the codebook of the linear block code.
	
	\begin{figure*}[!t]
		\begin{equation}
			\label{lapp}
			\begin{aligned}
				L_i^{\mathrm{APP,Y}} = \log \frac{\sum_{c_i=0, c^N \in \mathcal{L}_C} Q_{C^N \mid Y^N}\left(c^N \mid y^N\right)+\left(Q^{*}_{\mathcal{U}}\left(y^N\right)-\sum_{c^N \in \mathcal{L}_C} Q_{C^N \mid Y^N}\left(c^N \mid y^N\right)\right) P_{C_i \mid Y_i}\left(0 \mid y_i\right)}{\sum_{c_i=1, c^N \in \mathcal{L}_C} Q_{C^N \mid Y^N}\left(c^N \mid y^N\right) + \left(Q^{*}_{\mathcal{U}}\left(y^N\right) - \sum_{c^N \in \mathcal{L}_C} Q_{C^N \mid Y^N}\left(c^N \mid y^N\right)\right) P_{C_i \mid Y_i}\left(1 \mid y_i\right)} \\
				\text{where } \mathcal{L}_C = \left\{c^N \in\{0,1\}^N\mid c^N=u^N \mathbb{F}^{\otimes \log _2 N}, \forall u^N \in \mathcal{L}\right\}\\
				\text{and } Q_{C^N \mid Y^N}\left(c^N \mid y^N\right) = Q_{U^N \mid Y^N}\left(u^N \mid y^N\right)\text{, for } c^N=u^N \mathbb{F}^{\otimes \log _2 N}
			\end{aligned}
		\end{equation}
	\end{figure*}
	
	\ac{BCJR} algorithm~\cite{bcjr} realizes optimal \ac{SISO} decoding for any linear code, but its complexity grows exponentially in the number of $N-K$ for block codes. Pyndiah's algorithm~\cite{pyndiah} and \ac{SO-SCL}~\cite{yuan} are two methods to realize \ac{SISO} decoder for polar-like codes in polynomial complexity.
	
	Pyndiah’s algorithm can apply to any linear block code and approximates $L^{\mathrm{APP}}_i$ through a candidate list $\mathcal{L}_C$ in \eqref{lapp} from \ac{SCL} decoding,
	\begin{equation}
		L^{\mathrm{APP,P}}_i=\left\{\begin{array}{l}
			+\beta, \text { if }\left\{c_i=1, c^N \in \mathcal{L}_C\right\}=\varnothing \\
			-\beta, \text { if }\left\{c_i=0, c^N \in \mathcal{L}_C\right\}=\varnothing \\
			\log \frac{\max _{c_i=0, c^N \in \mathcal{L}_C} P_{C^N \mid Y^N}\left(c^N \mid y^N\right)}{\max _{c_i=1, c^N \in \mathcal{L}_C} P_{C^N \mid Y^N}\left(c^N \mid y^N\right)}, \text { otherwise.}
		\end{array}\right.
	\end{equation}
	In practical applications, a weighting factor $\alpha$ is multiplied to the extrinsic \ac{LLR} $L_i^{\mathrm{E,P}}$ to reduce the influence of approximation. The factor $\alpha$ and the saturation value $\beta > 0$ require optimization via practical trials.
	
	\Ac{SO-SCL} is designed for polar-like code. It fully utilizes the information of \ac{SCL} decoder to provide a more precise approximation. Firstly, \ac{SO-SCL} decoder approximates the sum of probabilities of all decoding paths in $\mathcal{U}$ and represents it as $Q^{*}_{\mathcal{U}}\left(y^N\right)$. Then $Q^{*}_{\mathcal{U}}\left(y^N\right)$ is adopted in soft output calculation~\eqref{lapp} to obtain a more accurate approximation. Similar to Pyndiah’s algorithm, a weighting factor $\alpha$ is adopted to the extrinsic \ac{LLR} in practical use. The computational complexity of \ac{SO-SCL} decoding is $\mathcal{O}\left(NL\log_2N\right)$ since it dose not conduct any operation with complexity higher than \ac{SCL} decoding.
	
	\subsection{BP Decoding for GLDPC-PC}
	
	\begin{figure}[!t]
		\centering
		\includegraphics[width=0.45\textwidth]{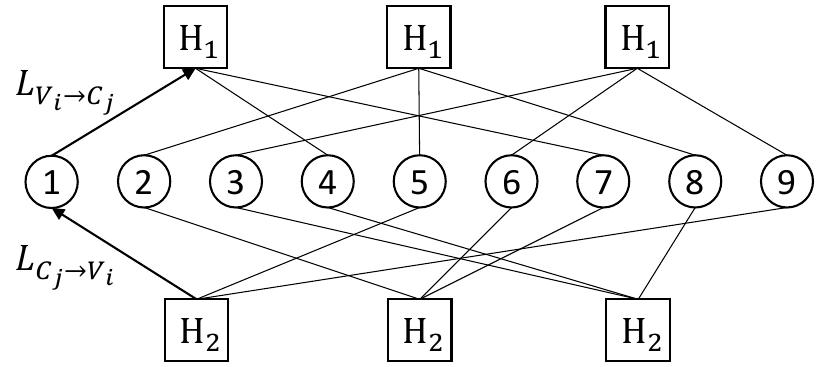}
		\caption{The Tanner graph of $\mathbf{\Gamma}_0$ with $\tilde{n}=3$. This graph has 9 \acp{VN} and 6 \acp{CN}. Every \ac{VN} is connected to two codes with parity check matrix $\mathbf{H}_1$ and $\mathbf{H}_2$.}
		\label{gldpc}
	\end{figure}
	
	The \ac{BP} decoding of \ac{GLDPC} codes is performed through iteratively passing extrinsic information between the \acp{VN} and \acp{CN} in the Tanner graph of $\mathbf{\Gamma}$~\cite{gldpc}. The Tanner graph of $\mathbf{\Gamma}_0$ with $\tilde{n}=3$ is shown in Fig.~\ref{gldpc}, where $\mathbf{H}_1$ and $\mathbf{H}_2$ are the parity check matrix of two different linear block codes with length $3$. Let $L_{V_i\rightarrow C_j},i\in \left[N\right],j\in \mathcal{N} \left(V_i\right)$ denote the extrinsic \acp{LLR} from \acp{VN} to \acp{CN} and $L_{C_j\rightarrow V_i},j\in \left[m\right],i\in \mathcal{N} \left(C_j\right)$ denote the extrinsic \acp{LLR} from \acp{CN} to \acp{VN}, where $\mathcal{N} \left(V_i\right)=\left\{j\mid \mathbf{\Gamma}_{i,j}=1\right\},\mathcal{N} \left(C_j\right)=\left\{i\mid \mathbf{\Gamma}_{i,j}=1\right\}$.
	
	The input of \ac{BP} decoder is $L^{\mathrm{in}}_i=L_i+L_i^A$ and the maximum number of iterations is set to $R$ in advance. In every iteration, each \ac{VN} $V_i,i\in\left[N\right]$ firstly sends its extrinsic information to its neighbor \acp{CN} via
	\begin{equation}
		L_{V_i\rightarrow C_j} = L_i^{\mathrm{in}} + \sum_{k \in \mathcal{N} \left(V_i\right), k\neq j} L_{C_k\rightarrow V_i},\quad \forall j \in \mathcal{N} \left(V_i\right).
	\end{equation}
	Afterwards, \acp{CN} possess \ac{SISO} decoding of their component codes and send the output extrinsic \acp{LLR} to corresponding \acp{VN}, i.e., $L_{C_j\rightarrow V_i}=L^{\mathrm{E}}_k$ if $V_i$ is the $k$-th \ac{VN} of $C_j$. Finally, the \ac{APP} \acp{LLR} is calculated by
	\begin{equation}
		L_i^{\mathrm{APP}} \approx L_i^{\mathrm{in}} + \sum_{k \in \mathcal{N} \left(V_i\right)} L_{C_k\rightarrow V_i},\quad\forall i\in \left[N\right].
	\end{equation}
	The \ac{APP} \acp{LLR} are only approximate values due to the cycles in $\mathbf{\Gamma}$. The decoder obtains $\hat{c}^N$ by hard decision based on $L_i^{\mathrm{APP}}$ and checks whether $\hat{c}^N$ is a codeword via whether the parity check of every component code is satisfied. The decoding terminates if $\hat{c}^N$ is a codeword or the number of iterations reaches $R$ and the $L_i^{\mathrm{APP}}$ at this time is the final soft output. 
	
	For \ac{BP} decoding, the \ac{SISO} decoder of polar-like component codes can adopt Pyndiah's algorithm or \ac{SO-SCL}. As described in~\cite{pyndiah}, the weighting factor $\alpha$ and the saturation value $\beta > 0$ need optimization for every iteration for Pyndiah's algorithm. By \ac{SO-SCL}, better performance is achieved even if $\alpha$ is constant, which is to be shown in simulations. 
	
	The \ac{BP} with \ac{SO-SCL} decoder realizes the \ac{SISO} decoding for \ac{GLDPC-PC} codes. The computational complexity of \ac{BP} with \ac{SO-SCL} is at most $\mathcal{O}\left(RmNL\log_2N\right)$, where $N$ is the largest block length of all the component codes. Whereas the complexity of \ac{SO-SCL} decoder for polar-like code is higher than the \ac{SISO} decoder for \ac{SPC}, the constraint of polar-like codes is more powerful than \ac{SPC}. Therefore, less iterations are required for \ac{GLDPC-PC} code to obtain similar performance with \ac{LDPC} code and the complexity increase can be compensated by this.
		
	\section{Iterative MIMO Detection and Decoding}
	\label{section4}
	
	This section presents a point-to-point \ac{MIMO} \ac{IDD} system coded by \ac{GLDPC-PC} codes. The details of the system are as follows.
	
	\subsection{System Model}
	Consider a point-to-point \ac{MIMO} system with $N_t$ transmit antennas and $N_r$ receive antennas. The $K$ bits message $s^K$ is encoded into $N$ bits codeword $c^N$. The systematic encoding of \ac{GLDPC} code is applied to obtain optimal performance in \ac{IDD}. 
	
	Consider the use of $16$-\ac{QAM} with Gray labeling so the input alphabet is $\mathcal{X}=\frac{1}{\sqrt{10}}\{\pm a \pm j b|a,b\in\{1,3\}\}$. A random interleaver permutes the encoded bits $c^N$ after the encoder and obtain $b^N$. The permuted codeword $b^N\in\{0,1\}^{N}$ are mapped into $T = N/N_t/M_c$ channel inputs $\boldsymbol{x}_t \in \mathcal{X}^{N_t}, t\in \left[T\right]$, where $M_c$ is the modulation order and $M_c=4$ for $16$-\ac{QAM}. Any $G = N_tM_c$ bits are mapped into a input vector $\boldsymbol{x}\in \mathcal{X}^{N_t}$ in a fixed way and let $\boldsymbol{x}=\operatorname{map}\left(b^{G}\right)$ denote this process. The total transmitted power is then obtained by $E_s = \mathop{\mathbb{E}}\left[\boldsymbol{x}^H\boldsymbol{x}\right]=N_t$. The transmitter model is shown in the upper area of Fig~\ref{mimo}.
	
	We consider the fading channel where the fading coefficients $\mathbf{H}_t\in \mathbb{C}^{N_r\times N_t}$ are different in every channel use. $\mathbf{H}_t$ is known perfectly to the receiver. For every channel use, the channel output is
	\begin{equation}
		\boldsymbol{y}_t = \mathbf{H}_t \boldsymbol{x}_t + \boldsymbol{n}_t,\quad t\in \left[T\right],
		\label{channel}
	\end{equation}
	where $\boldsymbol{y}_t \in \mathbb{C}^{N_r}$ is the received signal vector and $\boldsymbol{n}_t\in\mathbb{C}^{N_r}$ is \ac{AWGN} vector whose entries follow \ac{i.i.d.} complex Gaussian distribution with zero-mean and variance $2\sigma^2$, i.e., $\mathcal{C N}\left(0,2 \sigma^2\right)$. The noise variance $2\sigma^2$ is known to the receiver by long term measurement.

	\subsection{Iterative Detection and Decoding}
	
	\begin{figure}[!t]
		\centering
		\includegraphics[width=0.45\textwidth]{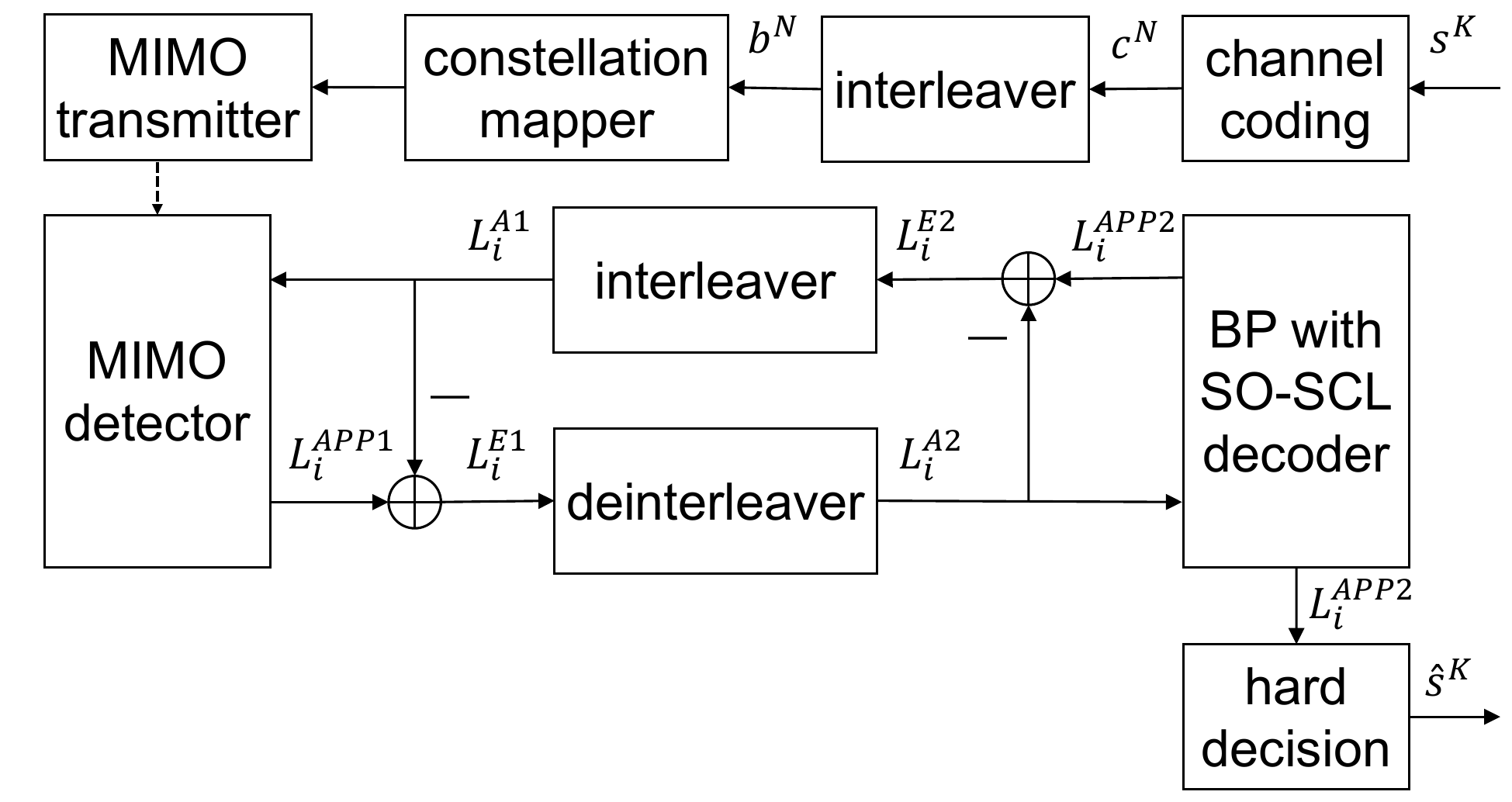}
		\caption{Transmitter and receiver scheme with iterative detection and decoding. In the receiver, the extrinsic \acp{LLR} are exchanged between MIMO detector and BP decoder.}
		\label{mimo}
	\end{figure}
	
	The \ac{IDD} is implemented via exchanging extrinsic \acp{LLR} between \ac{MIMO} detector and \ac{SISO} decoder of channel coding~\cite{idd}. The receiver model is shown in Fig.~\ref{mimo}. 
	
	The exchange starts from the \ac{MIMO} detector.
	The \ac{MIMO} detector treats a-priori \acp{LLR} $L_i^{\mathrm{A1}}$ and received vector $\boldsymbol{y}_t$ as input and outputs \ac{APP} \acp{LLR} $L_i^{\mathrm{APP1}}$. $L_i^{\mathrm{A1}}$ is all zero in the beginning. If $b_i$ is mapped into input vector $\boldsymbol{x}_t$ and is the $i_G$-th bit in $\boldsymbol{x}_t$, from~\cite{idd} we have
	
	\begin{equation}
		\begin{aligned}
			L_i^{\mathrm{A1}} & \triangleq \log \frac{P_{B_i}(0)}{P_{B_i}(1)},\\
			L^{\mathrm{APP1}}_i & \triangleq \log \frac{P_{B_i \mid Y^N}\left(0 \mid y^N\right)}{P_{B_i \mid Y^N}\left(1 \mid y^N\right)} \\
			& =\log \frac{\sum_{\boldsymbol{x}\in \mathcal{X}_i^0} p_{\mathbf{Y}\mid \mathbf{X}}\left(\boldsymbol{y}_t \mid \boldsymbol{x} \right) \exp\left(\sum_{j\in \mathcal{J}_{\boldsymbol{x}}}L_{j}^{\mathrm{A1}}\right)}{\sum_{\boldsymbol{x}\in \mathcal{X}_i^1} p_{\mathbf{Y}\mid \mathbf{X}}\left(\boldsymbol{y}_t \mid \boldsymbol{x} \right)\exp\left(\sum_{j\in \mathcal{J}_{\boldsymbol{x}}}L_{j}^{\mathrm{A1}}\right)},
		\end{aligned}
		\label{detector}
	\end{equation}
	where $\mathcal{X}_i^0 = \left\{ \boldsymbol{x}=\operatorname{map}\left(d^{G}\right)\mid d^{G} \in \{0,1\}^{G}, d_{i_G}=0 \right\}$,  $\mathcal{X}_i^1 = \left\{ \boldsymbol{x}=\operatorname{map}\left(d^{G}\right)\mid d^{G} \in \{0,1\}^{G}, d_{i_G}=1 \right\}$ and 
	$\mathcal{J}_{\boldsymbol{x}} = \left\{j\mid \operatorname{map}\left(d^G\right)=\boldsymbol{x}, d_j=0 \right\}$. From~\eqref{channel},
	\begin{equation}
		p_{\mathbf{Y}\mid \mathbf{X}}\left(\boldsymbol{y} \mid \boldsymbol{x} \right)=\frac{1}{\left(2 \pi \sigma^2\right)^N} 
		\exp\left(-\frac{1}{2 \sigma^2} \|\boldsymbol{y}-\mathbf{H}_t \boldsymbol{x}\|^2\right).
	\end{equation}
	
	The input of \ac{SISO} decoder is the result after deinterleaving the extrinsic information of \ac{MIMO} detector $L_i^{\mathrm{E1}}$. The \ac{APP} \acp{LLR} $L_i^{\mathrm{APP2}}$ is output and the extrinsic \acp{LLR} $L_i^{\mathrm{E2}}$ is interleaved and treated as one of the inputs of \ac{MIMO} detector. $L_i^{\mathrm{APP2}}$ and $L_i^{\mathrm{E2}}$ are defined in~\eqref{decoder} and computed by the \ac{BP} with \ac{SO-SCL} decoder.
	
	As shown in Fig.~\ref{mimo}, conventional extrinsic information exchange is achieved by
	\begin{equation}
		\begin{aligned}
			L_i^{\mathrm{E1}} & = L_i^{\mathrm{APP1}}-L_i^{\mathrm{A1}},\\
			L_i^{\mathrm{E2}} & = L_i^{\mathrm{APP2}}-L_i^{\mathrm{A2}}.
		\end{aligned}
	\end{equation}
	However, setting $L_i^{\mathrm{E2}} = L_i^{\mathrm{APP2}}$ can improve the performance for the \ac{MMSE}-\ac{PIC} detector adopted in this system~\cite{le-lapp}. Furthermore, we set $L_i^{\mathrm{E2}} = \rho L_i^{\mathrm{APP2}},\rho \in \left.\left(0,1\right.\right]$ to decrease the impact of this action.
	
	Let $R$ denote the total number of decoding iterations of \ac{BP}. Let $D$ denote the times of \ac{MIMO} detection. From the process of \ac{IDD} system, we can know that the \ac{BP} decoder performs $N_I=R/D$ rounds of iteration and exchanges extrinsic \acp{LLR} once.
	
	\subsection{SISO MMSE-PIC Detection}
	The computation of the \acp{LLR} in~\eqref{detector} entails enormous computational complexity, which grows exponentially in the number of $N_tM_c$. Therefore, several sub-optimal algorithms are proposed to approximate~\eqref{detector} in polynomial complexity. The low-complexity \ac{SISO} \ac{MMSE}-\ac{PIC} algorithm proposed in~\cite{mmse-pic} is considered in this system.
	
	For every input vector $\boldsymbol{x}_t$, received vector $\boldsymbol{y}_t$, a-priori \acp{LLR} $L_{i,j}^{\mathrm{A2}},i\in\left[N_t\right],j\in\left[M_c\right]$ of the bits in $\boldsymbol{x}_t$ and channel $\mathbf{H}_t$, \ac{MMSE}-\ac{PIC} detection outputs the \acp{LLR} of the $G$ bits in $\boldsymbol{x}_t$. It is summarized in the following five steps.
	
	\paragraph{Computation of Soft-Symbols}
	Every input vector $\boldsymbol{x}_t$ possesses $N_t$ symbols $x_i\in \mathcal{X},i\in \left[N_t\right]$ and every symbol possesses $M_c$ bits $b_{i,j}\in\{0,1\},j\in\left[M_c\right]$. $L_{i,j}^{\mathrm{A2}}$ denotes the a-priori information of the $j$-th bit in symbol $x_i$. Based on the a-priori information, the algorithm firstly computes the expectation and variance of every symbol $x_i$ by
	\begin{equation}
		\begin{aligned}
			\hat{x}_i &= \mathop{\mathbb{E}}\left[x_i\right] = \sum_{x\in\mathcal{X}} xP(x_i = x)\\
			E_i &=\operatorname{Var} \left[x_i\right] = \mathop{\mathbb{E}}\left[\left|e_i\right|^2\right]
		\end{aligned}
	\end{equation}
	where $e_i = x_i-\hat{x}_i$ and $P(x_i=x)=\prod_{j=1}^{M_c}P(b_{i,j}=d_j)$ if the $j$-th bit in $x$ is $d_j$. \Ac{LLR} can be converted into probability by
	\begin{equation}
		P(b_{i,j}=d) = \frac{1}{1+\exp\left((2d-1)L_{i,j}^A\right)}.
	\end{equation}
	
	\paragraph{Parallel Interference Cancellation}
	The algorithm then cancels the interference for every symbol $x_i$ induced by other symbols $x_j,j\neq i$ via
	\begin{equation}
		\hat{\boldsymbol{y}}_i=\boldsymbol{y}_t-\sum_{j, j \neq i} \mathbf{h}_j \hat{x}_j=\mathbf{h}_i x_i+\tilde{\mathbf{n}}_i
	\end{equation}
	where $\mathbf{h}_j$ is the $j$-th column of $\mathbf{H}_t$ and $\tilde{\mathbf{n}}_i=\sum_{j, j \neq i} \mathbf{h}_j e_j+\mathbf{n}$ corresponds to the \ac{NPI}.
	
	\paragraph{MMSE Filter-Vector Computation}
	All the \ac{MMSE} filter vectors are computed by
	\begin{equation}
		\mathbf{W}^H = \mathbf{A}^{-1}\mathbf{H}_t^H,
	\end{equation}
	where $\mathbf{A}=\mathbf{H}_t^H \mathbf{H}_t \boldsymbol{\Lambda}+N_0 \mathbf{I}_{N_t}$ with $\boldsymbol{\Lambda}$ denoting a $N_t\times N_t$ diagonal matrix having $\Lambda_{i,i}=E_i,i\in\left[N_t\right]$ and $N_0=2\sigma^2$. Every row $\mathbf{w}_i^H$ of $\mathbf{W}^H$ corresponds to the \ac{MMSE} filter vectors for symbol $x_i$.
	
	\paragraph{MMSE Filtering}
	The \ac{MMSE} filter vector $\mathbf{w}_i^H$ are then applied to obtain the estimate of symbol $x_i$ via
	\begin{equation}
		\tilde{x}_i = \frac{1}{\mu_i} \mathbf{w}_i^H \hat{\boldsymbol{y}}_i,
	\end{equation}
	where $\mu_i = \mathbf{w}_i^H \mathbf{h}_i$ denotes the bias. The \ac{NPI} variance $\nu_i$ can be computed as
	\begin{equation}
		\nu_i = \frac{1}{\mu_i} - E_i.
	\end{equation}
	
	\paragraph{LLR Computation}
	With the estimates and \ac{NPI} variances of all symbols, the extrinsic \acp{LLR} are calculated as
	\begin{equation}
		L_{i, j}^{\mathrm{E2}}=\frac{1}{\nu_i}\left(-\min _{a \in \mathcal{Z}_j^0}\left|\tilde{x}_i-a\right|^2+\min _{a \in \mathcal{Z}_j^1}\left|\tilde{x}_i-a\right|^2\right) .
	\end{equation}
	where $\mathcal{Z}^b_j$ denotes the set of symbols in $\mathcal{X}$ which the $j$-th bit is equal to $b$.
	
	\section{Numerical Results}
	\label{section5}
	
	This section provides the \ac{MC} simulations for the \ac{GLDPC-PC} under \ac{AWGN} channel and \ac{MIMO} \ac{IDD} system. The complexity of \ac{BP} decoding for \ac{GLDPC-PC} and \ac{LDPC} is balanced by reducing the number of iterations for GLDPC codes. 

	\subsection{AWGN Channel}
	
	\begin{figure}[!t]
		\centering
		\includegraphics[width=0.4\textwidth]{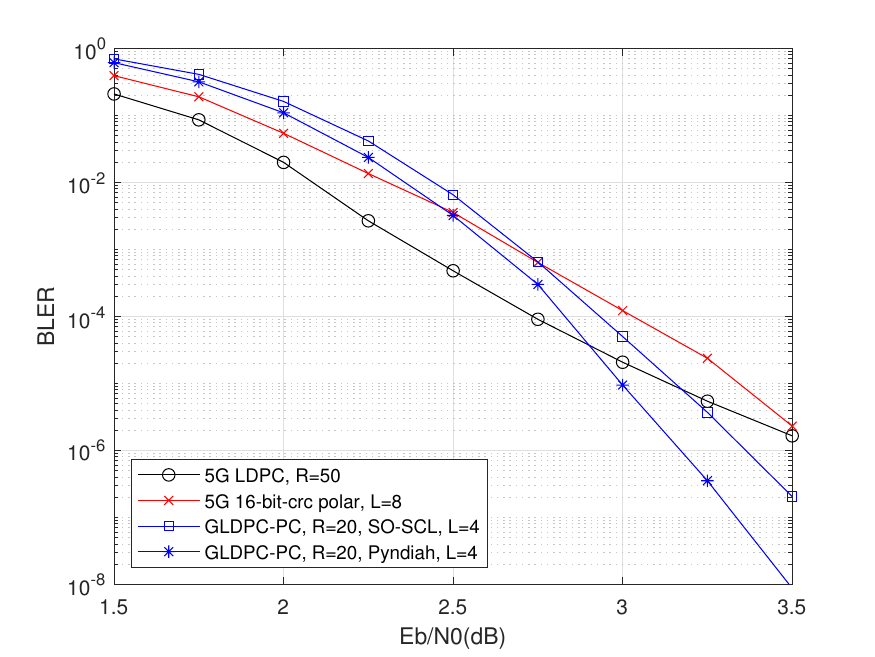}
		\caption{BLER performance of GLDPC-PC code, 5G LDPC code and 5G polar code under AWGN channel. The block length $N$ is 1024 and message bit length $K$ is 643.}
		\label{awgn}
	\end{figure}

	Fig~\ref{awgn} compares the (1024, 643) \ac{GLDPC-PC} code in Section~\ref{pcgldpc} under \ac{BP} decoding with Pyndiah's algorithm and \ac{SO-SCL}. The $\alpha$ for \ac{SO-SCL} decoder is 0.6. The $\alpha$ and $\beta$ for Pyndiah's algorithm are optimized for every iteration and $E_b/N_0$, so they are omitted here. The iteration round $R$ is 20 and list size $L$ for the both algorithms is 4. \Ac{SO-SCL} has better performance and is convenient to use, so in the following simulations we only consider \ac{SO-SCL}.
	
	Additionally, the (1024, 643) \ac{LDPC} code and polar code in \ac{5G} are compared with the \ac{GLDPC-PC} code. For \ac{LDPC} code, \ac{BP} decoding is adopted and maximum number $R$ of iterations is set to $50$. For polar code, a 16 bit \ac{CRC} code is concatenated to the \ac{5G} polar code and \ac{SCL} decoding is applied with list size $L=8$.
	
	Simulation results show that the \ac{GLDPC-PC} code performs better than \ac{LDPC} code below the \ac{BLER} $\approx 3\times 10^{-5}$ with \ac{SO-SCL}. Because of the irregular property~\cite{irregular} of \ac{5G} \ac{LDPC} code, the performance of it in low $E_b/N_0$ exceeds \ac{GLDPC-PC} code. \ac{GLDPC-PC} code with irregularity has the potential to perform better in this region.
	
	\subsection{MIMO Channel}

	\begin{figure}[!t]
		\centering
		\includegraphics[width=0.35\textwidth]{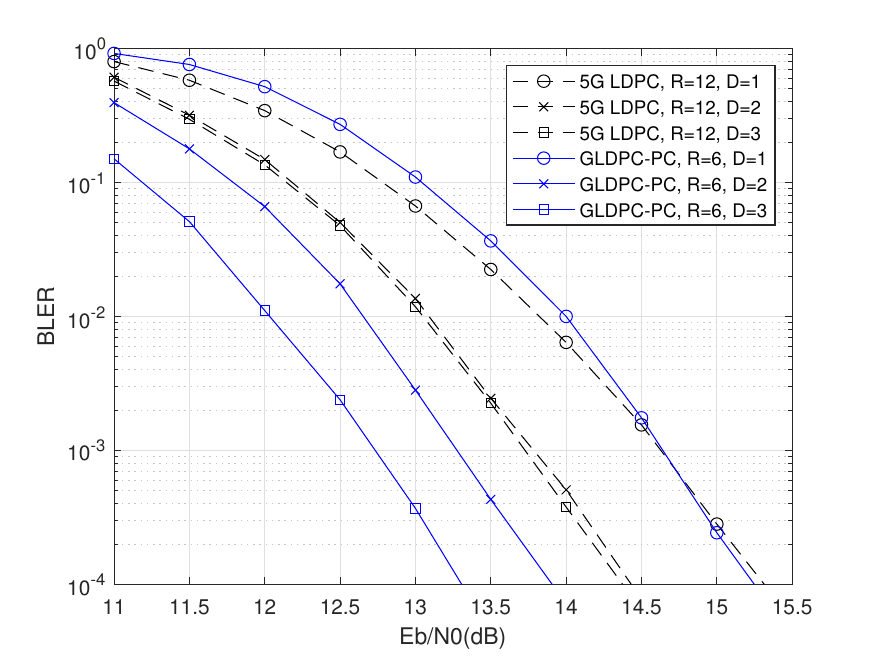}
		\caption{BLER performance for (1024, 643) GLDPC-PC code and \ac{5G} \ac{LDPC} code in $4\times 4$ MIMO channel. $N_I = R/D$ rounds of BP is performed after once detection.}
		\label{gldpc4}
	\end{figure}

	\begin{figure}[!t]
	\centering
	\vspace{-0.4cm}
	\includegraphics[width=0.35\textwidth]{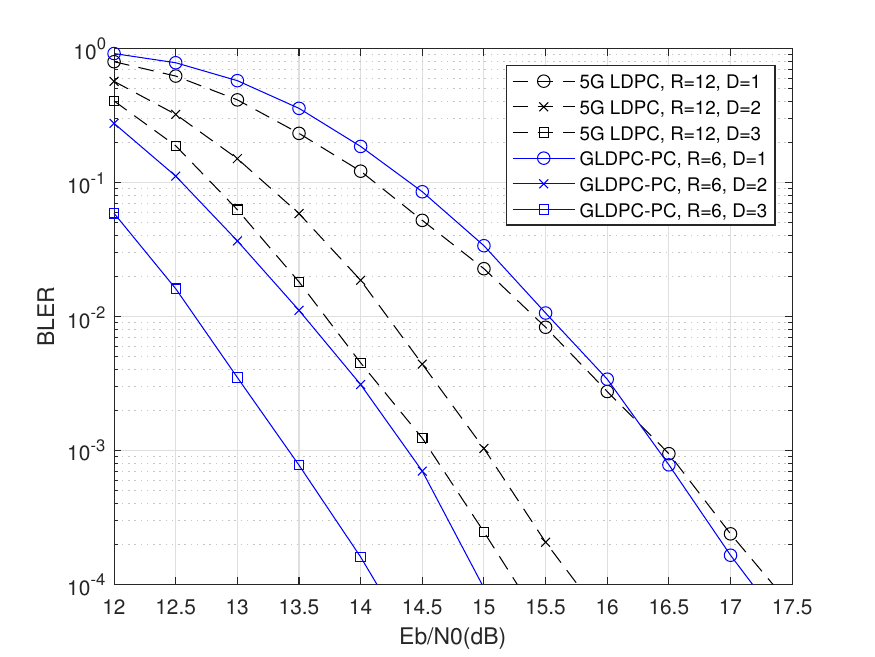}
	\caption{BLER performance for (1024, 643) GLDPC-PC code and \ac{5G} \ac{LDPC} code in $8\times 8$ MIMO channel. $N_I = R/D$ rounds of BP is performed after once detection.}
	\label{gldpc8}
	\end{figure}

	 Rayleigh fading channel is considered in all the simulations, e.g., the entries of $\mathbf{H}_t$ are \ac{i.i.d.} as $\mathcal{C N}\left(0,1\right)$ and $\mathbf{H}_{t},t\in\left[T\right]$ are independent to each other. The relation of $E_b/N_0$ and $E_s/N_0$ follows~\cite{idd}
	 \begin{equation}
	 	\left.\frac{E_b}{N_0}\right|_{\mathrm{dB}}=\left.\frac{E_s}{N_0}\right|_{\mathrm{dB}}+10 \log _{10} \frac{N N_r}{K N_t M_c} .
	 \end{equation}

	We adopt the (1024, 643) \ac{GLDPC-PC} code in Section~\ref{pcgldpc} and compare it  with the (1024, 643) \ac{5G} \ac{LDPC} code. The number of antennas is set to $N_t=N_r=4$ or $N_t=N_r=8$ and $16$-\ac{QAM} is considered.  $R$ is set to $12$ for \ac{LDPC} code and $6$ for \ac{GLDPC-PC} code. $R$ is consistent for all $D$ for the fairness among different values of $D$. \Ac{SO-SCL} decoder is adopted with list size $L=4$ and $\alpha = 0.6$ in \ac{BP} decoding. The factor $\rho$ is optimized for \ac{LDPC} and is set to $0.6$ for both \ac{LDPC} and \ac{GLDPC}.
	
	As shown in Fig.~\ref{gldpc4} and Fig.~\ref{gldpc8}, the \ac{GLDPC-PC} code performs better than \ac{LDPC} when $D=1$ and \ac{BLER} is low, which is consistent with the results in \ac{AWGN}. Furthermore, \ac{GLDPC-PC}  outperforms \ac{LDPC} when $D=2$ and $3$. At \ac{BLER} $\approx 10^{-4}$ and $D=3$, \ac{GLDPC-PC} outperforms \ac{LDPC} by about $1.1$~dB for both $4\times 4$ \ac{MIMO} and $8\times 8$ \ac{MIMO}.
	
	\section{Conclusion}
	\label{section6}
	
	In this work, we first introduce the \ac{GLDPC-PC} codes. These codes exhibit significantly lower error floors, which brings performance gains in the low \ac{BLER} region in \ac{AWGN} channel. The BP decoder is introduced for \ac{GLDPC-PC} codes, which adopts the low-complexity \ac{SO-SCL} decoder for polar-like codes. We also adapt this \ac{BP} with \ac{SO-SCL} decoder in \ac{MIMO} \ac{IDD} systems coded by \ac{GLDPC-PC} codes. Simulation results show the performance advantages of \ac{GLDPC-PC} code with fewer rounds of decoding iteration. The \ac{GLDPC-PC} code has the potential to perform better and exceed \ac{LDPC} code with more elaborate construction.

\end{document}